# Solving reviewer assignment problem in software peer review: An approach based on preference matrix and asymmetric TSP model

Yanqing Wang[*], Yu Jiang, Xiaolei Wang, Siyu Zhang, Yaowen Liang

School of Management, Harbin Institute of Technology

**Abstract:** Optimized reviewer assignment can effectively utilize limited intellectual resources and significantly assure review quality in various scenarios such as paper selection in conference or journal, proposal selection in funding agencies and so on. However, little research on reviewer assignment of software peer review has been found. In this study, an optimization approach is proposed based on students' preference matrix and the model of asymmetric traveling salesman problem (ATSP). Due to the most critical role of rule matrix in this approach, we conduct a questionnaire to obtain students' preference matrixes and convert them to rule matrixes. With the help of software ILOG CPLEX, the approach is accomplished by controlling the exit criterion of ATSP model. The comparative study shows that the assignment strategies with both reviewers' preference matrix and authors' preference matrix get better performance than the random assignment. Especially, it is found that the performance is just a little better than that of random assignment when the reviewers' and authors' preference matrixes are merged. In other words, the majority of students have a strong wish of harmonious development even though high-level students are not willing to do that.

## 1. Introduction

Peer review becomes increasingly essential to many people such as conference organizers, journal editors, grant administrators and educators (Cook et al. 2005; Xu et al. 2010) because qualified reviewers, especially high expertise reviewers, are always relatively limited intellectual resources. Sun et al. (2008) proposed that experts with high-level expertise would make useful and professional judgments on the project to be selected. Chen and Fan (2011) built up a model for measuring the match degree of reviewer's research discipline and that of proposal. The matching degree between reviewers and assigned proposal or paper manuscript determines the quality of peer review greatly (Xu et al. 2010) and improves the overall productivity (Karimzadehgan & Zhai 2012). Therefore, the most challenging issue of peer review, reviewer assignment problem (RAP), needs to be well solved urgently (Tsang 2013).

[*] Corresponding author. Tel.: +86 15124500998.
E-mail address: yanqing@hit.edu.cn (Y. Q. Wang)
Postal address: No.92, Xidazhi Street, Nangang District, Harbin 150001, China

In software industry, the concept of peer review derived from code inspection proposed by Fagan (1976) at IBM company. Afterward, many scholars and educators have proved the reliability and effectiveness of software peer review (also named as *peer code review*) in software industry (Meyer 2008; Devito Da Cunha & Greathead 2007) and software education field (Turner 2009; Li 2006; Li 2007). The strategy of reviewer assignment in software peer review has drawn attention of some researchers such as Turner (2009) and Li (2007), whereas they were still applying random assignment strategy or suggesting to use it in their research.

We have launched the research on software peer review since 2004. An information system dedicated to software peer review, EduPCR, was developed and has been updated for several times. Due to EduPCR is in an educational context, every student should play the roles of both reviewer and author equally, as shown in Fig. 1. Some contributions were made such as quality assurance (Wang et al. 2007), participants' behavior analysis (Wang et al. 2008), learning outcome analysis (Wang et al. 2011), assessment approach (Wang et al. 2012) and so on.

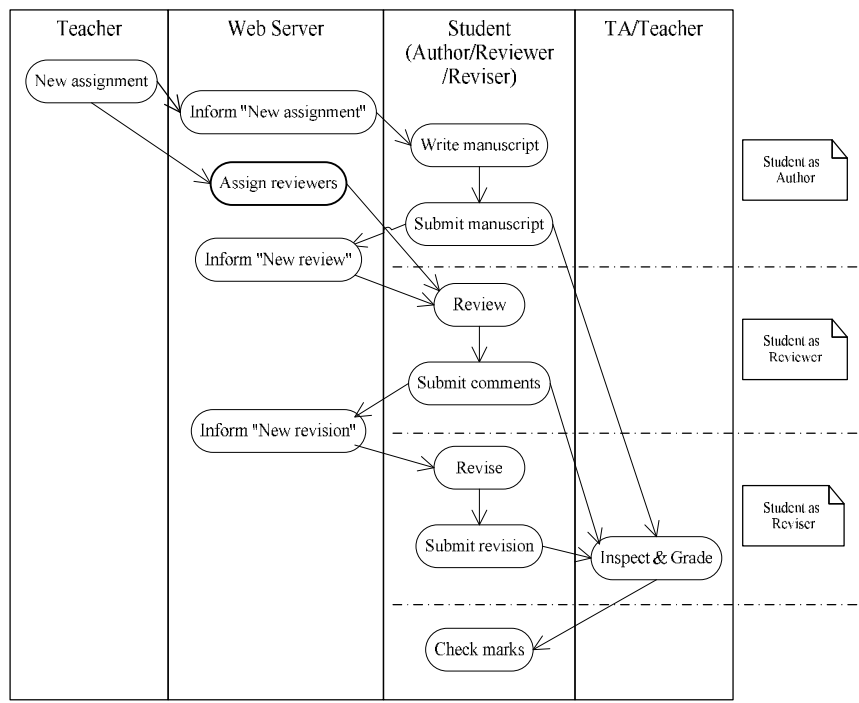

Fig. 1. Activity diagram of software peer review process in EduPCR

In Fig. 1, *author*, *reviewer* and *reviser* are three roles that every student plays in different stages; *manuscript* stands for the first version of a program written by an author according to the task set by teacher; *comments* indicate the suggestions to an author proposed by a reviewer; *revision* presents the revised version of an author's program. The web server informs students by a short message gateway.

So far, EduPCR has been adopting a random reviewer assignment strategy. While it is objective and likely fair, it is not quite satisfactory. The difference of students' competence on programming exists naturally, which can be measured by levels such as *high*, *middle*, and *low*. The assignments between the students with different competence levels will definitely produce different performance. The assignment between reviewer and author is conducted randomly in EduPCR so that we cannot effectively utilize the difference of students' competence and improve the overall learning outcome. Also, we do not know how students wish to be assigned and whether high-level students are willing to help low-level students achieve a *harmonious development*, in which all levels of students make approximately equal progress in a period of time.

## 2. Methodology

Integer linear programming is a very successful and popular method being used to solve RAP (Cook et al. 2005; Sun et al. 2008; Karimzadehgan & Zhai 2012). By referencing to the previous work, we recommend a novel approach to solve this problem with integer programming. The key issue is how to define the distance of each review pair, in format of {reviewer→author}, in the approach. We believe that the assignments between the students with different competence levels will achieve different learning outcome, so the assignment between a level of reviewers to a level of authors is defined as a distance. When the total distance is minimized, the maximum learning outcome is obtained.

### *2.1 Ranking students' competence*

The premise of solving the RAP in this study is to rank students' competence of programming. To make the level of student's competence more objective and practical, the accumulative average scores of students are taken as the input of computation. That is to say, the more tasks a student has finished, the more accurate this value is.

The detailed algorithm is not complicated. At the beginning, initialize every student's competence with *middle* level. When each program task is done, recalculate every student's updated average score. Afterward, cluster all students according to their competence of programming. The hierarchical clustering module in some popular software packages such as SPSS can handle it well. Without loss of generality, the students' competence of programming is ranked as three levels including *high*, *middle*, and *low*.

### *2.2 Defining rule matrix*

Having ranked students' competence, what we need to consider is the distance between every two levels rather than the competence distance of every two individual students. We name the distance square matrix $R_{3\times3}$ as rule matrix, as depicted in Formula 1. Rule matrix is actually a compact format of the distance matrix of all students.

$$R=\begin{bmatrix} a_{11} & a_{12} & a_{13} \\ a_{21} & a_{22} & a_{23} \\ a_{31} & a_{32} & a_{33} \end{bmatrix} \quad (1)$$

In Formula 1, the three lines represent reviewers' three levels including *high*, *middle*, and *low*. Similarly, the three columns stand for authors' three levels in the same order. Due to the essence of distance in a minimized objective, the smaller an item in *R* is, the higher priority the assignment between the corresponding reviewer and author will obtain. The item values in rule matrix are critical because they determine the results of integer programming.

*2.3 Forming distance matrix*

Assume that the number of students is *n* and all numbers build up a set *N* ($N = \{1...n\}$). Every student plays the roles of both reviewer and author, so the distance matrix is a square matrix, named with $D_{n\times n}$. The row number *i* ($i \in N$) of *D* represents the reviewer *i* while the column number *j* ($j \in N$) stands for author *j*.

It is easy to obtain the distance matrix *D* by expanding the rule matrix *R*. For example, if the ranking of reviewer *i* is *middle* level and that of author *j* is *high* level, then the element $d_{ij}$ in *D* is equal to $a_{21}$ (see Formula 1 for details). Of course, the elements on the diagonal line $d_{ii}$ are all set to positive infinity.

*2.4 Controlling the exit criterion of ATSP model*

Being in the context of education, every student has equal right to play dual roles of reviewer and author. Therefore, the RAP in this study is actually a *balanced assignment problem*. However, if we apply the model of traditional *balanced assignment problem*, we cannot control the size of subtours. In some RAPs such as that in software peer review, the control to the size of the shortest subtour is often required (Li 2007; Wang et al. 2012).

To the best of our knowledge, asymmetric traveling salesman problem (ATSP) has a close relationship with assignment problem (Martello & Toth 1987). Dai and Mao (1992) proved the relationship between assignment and shortest path problem and proposed to use shortest path problem to solve assignment problem. Frieze et al. (1992) considered the probabilistic relationship between the value of a random ATSP and the value of its assignment relaxation. The following integer linear programming formulation of the ATSP is well known as shown in Formula 2.

In the model, the definitions of *n*, *N*, $d_{ij}$ and *D* have the same meanings as mentioned in Section 2.3. The decision variable $X_{n\times n}$ is called assignment matrix, in which $x_{ij}$ will be 1 if reviewer *i* is assigned to author *j* or it will be 0 if not. The "subtour elimination" constraint in the third line is used to remove subtours from the consideration of preventing disjoint loops from occurring, in which *S* stands for any nonempty proper subset of *N* and |*S*| is the size of set *S*.

$$\min T = \sum_{i=1}^{n}\sum_{j=1}^{n} d_{ij} \cdot x_{ij}$$

$$s.t. \begin{cases} \sum_{j=1}^{n} x_{ij} = 1, \forall i \in N \\ \sum_{i=1}^{n} x_{ij} = 1, \forall j \in N \\ \sum_{i \in S}\sum_{j \in S} x_{ij} \leq |S| - 1, \forall S \subset N \\ x_{ij} \in \{0,1\} \end{cases} \quad (2)$$

However, the exit criterion of ATSP is too strict for the RAP in our study. In other words, we need not to get a single tour in this research. When every subtour is big enough, we stop iteration and get a better solution than that of ATSP. Therefore, we plan to apply ATSP model and make sure the size of every subtour is greater than or equal to a certain number by controlling the exit criterion of the model. Theoretically, the certain number could be an integer in the range of 2 to n/2 according to the real application requirement, whereas the constraint with a too small number, such as *two*, is likely to improve the negative impact of the "mutual admiration societies" pitfall mentioned by Li (2006).

## 3. Preference matrix

Undoubtedly, the above-defined rule matrix can manipulate the optimization results, but the emerging question is what students think or which assignment style they like most. In order to discover the students' preference to reviewer assignment, a questionnaire was conducted among the users of EduPCR, the undergraduate students in Year 1 through Year 4 majoring in Information Management and System in Harbin Institute of Technology in China. In total, 104 copies were delivered and 94 valid copies were received. There were about twenty questions in the questionnaire, in which the following two were relevant to this study:

*Q1. As a reviewer, which degree of student do you review can help you more with your learning outcome? (Please sort them in DESCENDING order):*

*A. the student with two levels superior to me;*

*B. the student with one level superior to me;*

*C. the student with the same level as me*

*D. the student with one level inferior to me;*

*E. the student with two levels inferior to me;*

*Order: ________*

*Q2. As an author, by which degree of student is your program reviewed can help you more with your learning come? (Please sort them in DESCENDING order):*

*A. by the student with two levels superior to me;*

*B. by the student with one level superior to me;*

*C. by the student with the same level as me*

*D. by the student with one level inferior to me;*

*E. by the student with two levels inferior to me;*

*Order:* _____________

The following stages constitute the algorithm of generating preference matrix and converting it to rule matrix.

(1) Initializing. Every value in the preference matrix is set to 0;

(2) Determining weights for options. When a student chooses an option to any one question mentioned above, the number of the affected items in preference matrix may be various. For example, an option "*A*" will affect item $a_{13}$ alone while an option "*C*" affects $a_{11}$, $a_{22}$ and $a_{33}$. To eliminate the influence of repeated computation, the weights to option "*A*" through "*E*" are set to 1, 1/2, 1/3, 1/2 and 1 respectively.

(3) Determining weights for positions. The appearance sequence of options determines their priorities. A prior chosen option implies that a student has stronger preference to it. Therefore, we adopt a common approach in statistics, the reciprocal of *e*, to set the weights for options in the first through the fifth position with $e^0, e^{-1}, e^{-2}, e^{-3}, e^{-4}$ respectively;

(4) Computing initial preference matrixes. We multiply option weight by position weight and accumulate the result value to the corresponding item in the preference matrix. For example, suppose that one student gives the answer to one question like "*B*...", two items are affected, i.e. $a_{21}=a_{21}+(1/2)e^0$, $a_{32}=a_{32}+(1/2)e^0$. The two questions are computed separately so that two preference matrixes $R_1$ and $R_2$ are obtained, the former is from the view of reviewer (part (a) in Table 1) and the latter is from the view of author (part (b) in Table 1).

Table 1 Initial preference matrixes of students

| | | author | | | | | reviewer | | |
|---|---|---|---|---|---|---|---|---|---|
| | | high | middle | low | | | high | middle | low |
| reviewer | high | 8.75 | 5.26 | 2.10 | author | high | 6.75 | 2.42 | 2.28 |
| | middle | 24.66 | 8.75 | 5.26 | | middle | 22.47 | 6.75 | 2.42 |
| | low | 35.92 | 24.66 | 8.75 | | low | 50.25 | 22.47 | 6.75 |

(a) $R_1$ by reviewer's preference  (b) $R_2$ by author's preference

(5) Normalizing and transposing. Normalization is easy to be finished by dividing every item

in a preference matrix by the least value in it. In addition, $R_1$ and $R_2$ come from different points of view by reviewer and author so that one of them must be transposed before they are merged. Thus, the normalized matrix of $R_1$ and the transposed and normalized matrix of $R_2$ are listed in Formula 3 as follows.

$$R_3 = \begin{bmatrix} 4.17 & 2.51 & 1.00 \\ 11.74 & 4.17 & 2.51 \\ 17.11 & 11.74 & 4.17 \end{bmatrix}, R_4 = \begin{bmatrix} 2.96 & 9.86 & 22.04 \\ 1.06 & 2.96 & 9.86 \\ 1.00 & 1.06 & 2.96 \end{bmatrix} \quad (3)$$

(6) Merging. As mentioned above, the peer review model is dedicated to education and every student should play the roles of both author and reviewer. Therefore, we mean to research $R_1$ and $R_2$ conjunctively to see the effect of integration preference after we study them separately. In Formula 4, $M$ is the least value in the matrix $(R_3 + R_4)$ and $R_5$ is the preference matrix after merging and normalization.

$$R_5 = (R_3 + R_4)/M = \begin{bmatrix} 1.00 & 1.79 & 3.38 \\ 1.75 & 1.00 & 1.79 \\ 2.46 & 1.75 & 1.00 \end{bmatrix} \quad (4)$$

From the preference matrix $R_5$, it is found that the students' preference pairs of reviewer assignment are listed in descending order as follows:
- assigning high-level reviewer to low-level author;
- assigning low-level reviewer to high-level author;
- assigning adjacently higher-level reviewer to author;
- assigning adjacently lower-level reviewer to author;
- assigning same-level reviewer to author.

(7) Reversing values. The biggest value in preference matrixes mentioned above implies that students want to be assigned in its corresponding assignment pair most. However, in the rule matrix applied in integer programming mentioned in Section 2, the smallest value has the highest programming priority. Therefore, to convert the preference matrixes to the rule matrixes, we reposition all values in the preference matrixes reversely, as shown in Formula 5.

$$R_3' = \begin{bmatrix} 4.17 & 11.74 & 17.11 \\ 2.51 & 4.17 & 11.74 \\ 1.00 & 2.51 & 4.17 \end{bmatrix}, R_4' = \begin{bmatrix} 2.96 & 1.06 & 1.00 \\ 9.86 & 2.96 & 1.06 \\ 22.04 & 9.86 & 2.96 \end{bmatrix}, R_5' = \begin{bmatrix} 3.38 & 1.79 & 1.00 \\ 2.46 & 3.38 & 1.79 \\ 1.75 & 2.46 & 3.38 \end{bmatrix} \quad (5)$$

## 4. Investigation and analysis

This study involved twenty three freshmen who had completed their preliminary programming class *C Programming* in fall semester of 2011. There were twelve programming tasks in total.

To make sense of how much the optimization approach with preference matrix is better than the random assignment strategy, a comparative study was performed. Because the course had been finished when we carried out the investigation, our research plan was to compare the performance of the actual random assignment with the simulated optimization assignment.

*4.1 Implementing the optimization approach*

The optimization approach was implemented in the following procedure.

(1) Taking students' accumulative average scores as input, their competence of programming was determined by the hierarchical clustering module of SPSS software so that students were classified into three levels including *high*, *middle* and *low*. In fact, the level of one particular student is dynamic because his/her average score may vary with tasks;

(2) Forming the distance matrix *D* by retrieving rule matrix *R*;

(3) With the assistance of the software ILOG CPLEX, the program was written in *Java* language. In order to make the following comparative study more accurate, the size of the shortest subtour was determined as *three* because the same number was utilized in the algorithm of random assignment to be mentioned in the following sub-section;

(4) During the integer programming process, the size of every subtour was checked after each iteration. When the size of each subtour was greater than or equal to *three*, the iteration was stopped and an optimal solution was obtained. Finally, the reviewer assignment result was acquired by parsing the decision variable matrix $X_{23 \times 23}$.

*4.2 Comparative study with random assignment*

By searching the data in the database of EduPCR system, the data of actual reviewer assignment applying random assignment strategy were extracted.

(1) Building up the random assignment matrix $A_{23 \times 23}$ based on the data just mentioned. $A_{23 \times 23}$ has an identical structure and meaning with decision variable $X_{23 \times 23}$ but has different values in it;

(2) Summarizing the total distance values separately. With the multiplication of two matrixes as in Formula 6, six total distance values $T_{r1}$, $T_{r2}$, $T_{r3}$, $T_{o1}$, $T_{o2}$ and $T_{o3}$ were obtained. $T_{r1}$, $T_{r2}$ and $T_{r3}$ mean the total distance values by random assignment while $T_{o1}$, $T_{o2}$ and $T_{o3}$ stand for the total distance values after the optimization. $D_1$, $D_2$ and $D_3$ were expanded with rule matrixes $R_3'$, $R_4'$ and $R_5'$ respectively.

$$\begin{aligned}
T_{r1} &= sumproduct(D_1, A), \quad T_{r2} = sumproduct(D_2, A), \quad T_{r3} = sumproduct(D_3, A) \\
T_{o1} &= sumproduct(D_1, X_1), \quad T_{o2} = sumproduct(D_2, X_2), \quad T_{o3} = sumproduct(D_3, X_3)
\end{aligned} \quad (6)$$

(3) Computing the relative ratio with the formula $C=T_r/T_o$. The measure is used to evaluate the optimization performance. The total distance by random assignment is numerator, so a bigger value indicates a better performance. Finally, the relative ratio values of each task and average were acquired, as shown in Table 2.

From the relative ratio values in Table 2, it can be found that the average values of $C_1$ and $C_2$ are both much greater than 1.0. That is to say, as long as we conduct the optimization using rule matrix $R_3{'}$ or $R_4{'}$ in Formula 5, we can achieve a better performance than that of random assignment. However, when we merge the reviewers' preference matrix and the authors' preference matrix, the final relative radio decreases to 1.11, which is somewhat disappointing.

Table 2 Comparison of random assignment to optimization assignment

| Task No. | $T_{o1}$ | $T_{r1}$ | $C_1$ | $T_{o2}$ | $T_{r2}$ | $C_2$ | $T_{o3}$ | $T_{r3}$ | $C_3$ |
|---|---|---|---|---|---|---|---|---|---|
| 1 | 101.82 | 119.55 | 1.17 | 73.08 | 88.08 | 1.21 | 62.68 | 67.70 | 1.08 |
| 2 | 95.91 | 145.00 | 1.51 | 68.08 | 127.32 | 1.87 | 44.83 | 57.17 | 1.28 |
| 3 | 101.82 | 162.64 | 1.60 | 73.08 | 173.96 | 2.38 | 44.65 | 50.37 | 1.13 |
| 4 | 95.91 | 148.56 | 1.55 | 68.08 | 128.88 | 1.89 | 44.24 | 55.25 | 1.25 |
| 5 | 101.82 | 119.55 | 1.17 | 73.08 | 88.08 | 1.21 | 67.70 | 67.70 | 1.00 |
| 6 | 101.82 | 139.09 | 1.37 | 73.08 | 122.32 | 1.67 | 54.66 | 59.68 | 1.09 |
| 7 | 101.82 | 127.12 | 1.25 | 73.08 | 107.04 | 1.46 | 62.19 | 64.49 | 1.04 |
| 8 | 93.48 | 113.33 | 1.21 | 67.16 | 92.24 | 1.37 | 61.32 | 60.31 | 0.98 |
| 9 | 95.91 | 142.95 | 1.49 | 68.08 | 134.44 | 1.97 | 55.67 | 58.18 | 1.05 |
| 10 | 101.82 | 125.46 | 1.23 | 73.08 | 93.08 | 1.27 | 65.19 | 65.19 | 1.00 |
| 11 | 95.91 | 168.55 | 1.76 | 68.08 | 178.96 | 2.63 | 35.83 | 47.86 | 1.34 |
| 12 | 93.48 | 125.00 | 1.34 | 67.16 | 96.96 | 1.44 | 56.01 | 60.10 | 1.07 |
| Avg. | 98.46 | 136.40 | *1.39* | 70.43 | 119.28 | *1.70* | 54.58 | 59.50 | *1.11* |

*4.3 Result analysis*

Why does this phenomenon happen? From the values in the first line of $R_1$, it is undoubted that high-level reviewers prefer to review the work by high-level authors rather than low-level ones. Similarly, from the values in the first line of $R_2$, it is obvious that high-level authors wish their work to be reviewed by high-level reviewers rather than low-level ones. There is no need to blame high-level students since their behavior is rational. However, in the merged preference matrix $R_3$, the top two preference pairs are {high→low} and {low→high}. It seems like a contradiction. After having deeply analyzed the values in $R_1$ and $R_2$, the reason is discovered. It is because the low-level students are much more eager to cooperate with high-level students than the high-level students are. With strong wish of *harmonious development*, the low-level students are trying to get more help by persuading teacher to assign them to the high-level authors and assign the high-level reviewers to them as well.

## 5. Conclusion and future work

RAP becomes a pressing concern due to the increasing trend of peer review practices by many people such as conference organizers, journal editors and grant administrators (Karimzadehgan & Zhai 2012). Similarly, in the field of software peer review, the research on reviewer assignment is needed urgently since it plays the roles of both assuring program's quality and enhancing learning outcome (Wang et al. 2012).

In the context of software peer review, an optimization approach is put forward. The key issues consist of obtaining students' preference matrix and controlling exit criterion of ATSP model. The highlight of this study is the research on the students' preference matrix based on questionnaire. The subsequent investigation shows the practical value of the optimization approach and the performance of the assignment with students' preference matrix is better than that of the random assignment.

Even though taking software peer review as an example, the contribution in this study can be applied to many other fields. For example, either with paper selection in journal (and conference) or with proposal selection in funding agencies, if we can rank reviewers and authors by their domain competence, the approach in this paper could be of great reference value.

However, RAP is definitely quite a challenging issue nowadays (Xu et al. 2010). In this study, we have just introduced an optimization approach and applied it to software educational filed. There is still much work to do. For example, since rule matrix can manipulate optimization results, can we construct it to achieve the goal of specialized education? That is to say, can we design a rule matrix for the benefit of low-level students alone or high-level students alone? As another example, if the rule matrix can be designed, the orientation effect of different rule matrixes on the competence improvement of organizations will be a very valuable topic in management science field.